\documentclass{PoS}

\usepackage{graphicx}
\usepackage{latexsym,amsmath,amssymb}

\usepackage[authoryear]{natbib}
\bibpunct{(}{)}{;}{a}{}{,} 

\title{The Local Dark Matter Density}

\ShortTitle{The Local Dark Matter Density}

\author{Fabrizio Nesti,$^a$ \speaker{Paolo Salucci}$^{b}$\vspace*{2ex}
\\
\llap{$^a$}University of L'aquila - I-67100, L'Aquila, Italy\\
\llap{$^b$}SISSA\\
E-mail: \email{nesti@aquila.infn.it}, \email{salucci@sissa.it}}

\abstract{We present the recent robust determination of the value of the Dark Matter density at the
  Sun's location ($\rho_\odot$) with a technique that does not rely on a global mass-modeling of the
  Galaxy.  The method is based on the local equation of centrifugal equilibrium and depends on local
  and quite well known quantities such as the angular Sun's velocity, the disk to dark contribution
  to the circular velocity at the Sun, and the thin stellar disk scale length. This determination is
  independent of the shape of the dark matter density profile, the knowledge of the rotation curve
  at any radius, and the very uncertain bulge/disk/dark-halo mass decomposition.  The result is:
  $\rho_\odot=0.43 (0.11)(0.10)\,$GeV/cm$^{3}$, where the quoted uncertainties are due to the
  uncertainty a) in the slope of the circular-velocity at the Sun location and b) in the ratio
  between this radius and the exponential length scale of the stellar disk.  The devised technique
  is also able to take into account any future improvement in the data relevant for the estimate.}

\FullConference{VIII International Workshop on the Dark Side of the Universe,\\
		June 10-15, 2012\\
		Rio de Janeiro, Brazil}

\begin{document}

\section{Introduction}

Galaxy rotation curves (e.g.~\citealt{rubin80,bosma81}) have unveiled the presence of a dark ``mass
component'' in spirals.  They are pillars of the paradigm of massive dark halos, composed of a still
undetected kind of matter surrounding the luminous part of galaxies. The kinematics of spirals shows
universal systematics~\citep{PSS,URC2}, which seems to be at variance with the predictions emerging
from simulations performed in the $\Lambda$ cold dark matter ($\Lambda$CDM) scenario,
(e.g.~\citealt{NFW96}), the current cosmological paradigm of galaxy formation
(e.g.~\citealt{gentile04}) (However, see \cite{Maccio':2011eh}).

Dedicated searches of DM particle candidates have intensified in recent years: direct-detection
experiments look for the scattering of DM particles inside the detectors, which clearly is
proportional to the DM density in the Sun's region $\rho_\odot\equiv \rho_{DM}(R_\odot)$.  Indirect-detection experiments
searching for the secondary particles (e.g. neutrinos) produced by DM annihilation at the center of
the Sun or Earth depend on the DM density inside these objects which in turn is driven, via a
capture mechanism, by the same local DM density~$\rho_\odot$.  Thus, in both direct
and indirect searches the knowledge  of the the local density $\rho_\odot$ is very important for a
precise prediction of the searched signal or to obtain reliable bounds on the DM particle
cross-section.

A value of $\rho_\odot =0.3\, {\rm GeV/cm}^{3}$ has been routinely quoted, but its origin is not
clear, neither supported by data (see the introduction of~\cite{SNG}). An example is the work by
\cite{caldwell81} that devised what can be considered as the standard method  to
determine the value of $\rho_{\odot}$ from a variety of  observations. Their
resulting value $0.23^{+0.23}_{-0.12}$GeV/cm$^{3}$  is however uncertain and moreover plagued by
very outdated kinematics.
 
In general, it is quite simple to infer the distribution of dark matter in {\it spiral
  galaxies}. Spiral's kinematics, in fact, reliably traces the underlying gravitational potential
\citep{PSS,URC2}. Then, from co-added and/or individual RCs, we can build suitable mass models that
include stellar and gaseous disks,  along with a spherical bulge and a dark halo. More in detail, by
carefully analyzing (high quality) circular velocity curves, with the help of relevant photometric
and HI data, one can derive the halo density at different radii.  The accuracy of the measurements
and of the analysis is excellent and the results obtained are at the heart of the present debates on
Galaxy formation (e.g.~\citealp{gentile04,gentile05,deblok09}; an innovative review on this issue
is found at {\tt www.sissa.it/ap/dmg/dmaw\_presentation.html})

To measure $\rho_\odot$, instead, is far from simple, because the MW kinematics, unlike that of
external galaxies, does not trace the gravitational potential straightforwardly.  We do not directly
measure the circular velocity of stars and gas but rather the terminal velocity $V_T$ of the
rotating HI disk, and this only inside the solar circle (e.g.~\citealt{McClure07}). This velocity is
related to the circular velocity $V(r)$, for $r<R_{\odot}$, by means of $V(r)= V_T(r)+ V_{\odot} \
r/R_{\odot}$, where $R_\odot\simeq 8$ kpc is the distance of the Sun from the Galaxy center and
$V_\odot$,  the value of the circular velocity at the Sun's position.  Both quantities are known
within an uncertainty of 5\% - 10\% (e.g.~\citealt{MB09}) which triggers a similar uncertainty in
the derived amplitude and slope of the circular velocity. As a result, (see~ also \citealt{sofue1}),
the MW circular velocity from 2\,kpc to 8\,kpc is known within a not negligible uncertainty:
 \begin{eqnarray}
\label{eq:Vcirc}
&&V(r)=(215 \pm 30) \ {\rm km/s}\\
&&d{\rm log} V(r)/d{\rm log}r  \equiv \alpha(r) = 0.0 \pm  0.15\,.
\label{eq:slope}
\end{eqnarray}

The variations in equation (\ref{eq:Vcirc}) are due to a mix of observational errors in the
kinematics, to  uncertainties in the values of $R_\odot$ and $V_\odot$ and to actual radial variations of
$V$.  The first two trigger also part of the possible range of the velocity slope (\ref{eq:slope}).
Data show that the radial variations of $\alpha(r)$ are, instead,  quite small: $d\alpha(r)/dr\simeq0\pm
0.03/$kpc $\simeq 0$; in the disk region the MW RC can be approximated by a straight line, whose
slope however is mildly uncertain.

The outer (out to 60 kpc) MW ``effective'' circular velocity $V(r)=(GM(r)/r)^{1/2}$ is even 
more uncertain and it depends on the assumptions made on dynamical and structural properties of its
estimators.  It appears to decline with radius, with quite an uncertain slope $ d{\rm log}
V(r)/d{\rm log} r =-0.20^{+0.05}_{-0.15}\qquad(R_\odot<r<60\,\text{kpc})$~\citep{battaglia05,xue08,brown09}.  These
uncertainties, combined with the intrinsic ``flatness'' of the RC in the region specified above
(that complicates the mass modeling even in the case of a high-quality RC~\citep{Tonini}), make it
very difficult to obtain a reliable bulge/disk/halo mass model and consequently an accurate estimate
of~$\rho_\odot$.

A different approach  was devised by~\cite{SNG}; the idea is to resort to the equation of
centrifugal equilibrium (see~\citealt{fall80} for details) and to use recent
results found in  external galaxies~\citep{URC2}. Let us start with 
\begin{equation}
V^2/r = a_H+a_D+a_B\,,
\label{eq:vtot}
\end{equation}
where $a_H$, $a_D$, and $a_B$ are the radial accelerations generated by the halo, stellar disk, and
bulge.  Taking first the (quite good) approximation of spherical DM halo, we have $a_H \propto
r^{-2}\int_0^r \rho_H(r) \ r^2 dr$. A similar relation holds for the bulge.  Therefore, by
differentiating equation~(\ref{eq:vtot}), we obtain the DM density at any radius in terms of the
local angular velocity $\omega(r)=V/r $, the RC slope $\alpha(r)$, the disk-to-dynamical mass ratio
$\beta(r)$ (see later), and the bulge density $\rho_B(r)$:
\begin{equation}
\rho_H (r) =   \omega(r)^2 [F_{tot}(\alpha(r)) - F_D (\beta(r))]-\rho_B(r).
\label{eq:rhoDM1}
\end{equation}
with $F_{tot}$ and $F_D$ known functions. 

In external spirals, equation~(\ref{eq:rhoDM1}) is not useful for determining the DM density at
different radius because 1) it collapses for $r<R_D$ where $F_{tot} \simeq F_D$ and where the bulge
density is dominating, $\rho_B\gg \rho_H$,  and 2) the radial variations of $\alpha(r)$ have
non-negligible observational uncertainties; 3) the quantity $\omega$ is known with far less accuracy
than $V$, the main kinematical observational quantity.

Instead, when estimating $\rho_\odot$, i.e.\ the density of the MW DM halo at a {\it specific}
radius (the Sun position), the above drawbacks disappear, infact: 1) we have  that $F_{tot}(R_\odot)\gg
F_D(R_\odot)$ because $R_\odot>3 R_D$, so equation~(\ref{eq:rhoDM1}) does not collapse and, as a
bonus, the most uncertain term of its r.h.s.\ turns out to be very small 2)
$\omega_\odot$ is very precisely measured; and 3) at the Sun's position the bulge density
$\rho_B(R_\odot)$ is totally negligible, $< \rho_H/50$ (e.g.~\citealt{sofue2}).   

The method is clearly simpler for a spherically symmetric DM halo, and if we assume an
infinitesimally thin disk for the distribution of stars in the Galaxy.  Nevertheless, in~\cite{SNG}
the effects of a possible halo oblateness and disk thickness were dealt with.  

Our claim \citep{SNG} is that by means of this method we to obtain an independent and reliable
determination of the local DM halo density and of its intrinsic uncertainty.

\medskip

\section{A new local method for determining $\rho_\odot$}

We model the Galaxy as composed by a stellar exponential thin disk plus an unspecified spherical DM
halo with density profile $\rho_H(r)$.  For the present work we can neglect the thick stellar and
the HI disk because their surface density, between $2\,$kpc and $R_\odot$, are 100 to 5 times
smaller than the stellar surface density~\citep{nakanishisofue}.  Similarly, we neglect the stellar
bulge because, as mentioned above, its spatial density at $R_\odot$ is virtually zero
(e.g.~\citealt{sofue2}). It is worth noticing that the standard method of galaxy modeling and its
variants cannot take these very simplifying assumptions because the global modeling involves all
these mass components in a crucial way.

Let us rewrite the equation of centrifugal equilibrium by subtracting the disk component from the
total acceleration. From its radial derivative we then find
\begin{equation}
\rho_H(r)=\frac{X_q}{4 \pi G r^2}\, \frac{d}{dr}\left[r^2\left(\frac{V^2(r)}{r}-a_{\rm D}(r)\right)\right],
\label{eq:rhohalo}
\end{equation}
where $X_q$ is a factor correcting the spherical Gauss law used above in case of oblateness $q$ of
the DM halo~\citep{SNG}.  The $X_q$ correction is very small: $X_q\simeq1.00$--$1.05$.

The disk component can be reliably modeled as a Freeman stellar exponential thin
disk of length scale $R_{\rm D}=(2.5\pm 0.2) \ {\rm
  kpc}$~\citep{PR04,juric08,robin08,reyle09}. The stellar surface density is then: $\Sigma(r) =
(M_{\rm D}/2\pi R_{\rm D}^2)\,e^{-r/R_{\rm D}}$.  Also, the disk can be considered infinitesimally
thin. In fact, its thickness $z_0$ is small, $z_0 \sim 250\,$pc \citep{juric08} and moreover
$z_0\ll R_D<R_\odot$.  We can thus write $a_{\rm D}(r)= \frac{GM_{\rm D}r}{R_{\rm
    D}3}(I_0 K_0 - I_1 K_1)\, X_{z_0}$, where $I_n$ and $K_n$ are the modified Bessel functions
computed at $r/2R_{\rm D}$, and $X_{z_0}\simeq0.95$ accounts for the nonzero disk thickness.

Since only the first derivative of the circular velocity $V(r)$ enters in (\ref{eq:rhohalo}) and in
any case this function in the solar neighborhood is almost linear, we can write
\begin{equation}
V(r)=V_{\odot} [1+ \alpha_\odot\, (r-R_\odot)/R_\odot ]\,,
\end{equation}
where $\alpha_\odot=\alpha(R_\odot)$ is the velocity slope at the Sun's radius. Then equation
(\ref{eq:rhohalo}) becomes 
\begin{equation}
\rho_H(r)= \frac{X_q}{4 \pi G}\left[\frac{V^2(r)}{r^2} (1+2 {\alpha_\odot})
  -\frac{GM_{\rm D}}{R_{\rm D}^3} H(r/R_{\rm D})\,X_{z_0}\right],
\label{rhoh}
\end{equation}
with $ 2H(r/R_{\rm D})= (3I_0 K_0 - I_1 K_1) + (r/R_ D)(I_1 K_0 - I_0 K_1)$.  Equation~(\ref{rhoh})
holds at any radius outside the bulge region and measures $\rho_H( R_\odot) \equiv \rho_\odot$ by
subtracting the density of the stellar component from the one of the whole gravitating matter.
 
The disk mass can be parametrized~\citep{PS90} by $M_{\rm D}= \beta \ 1.1 \ G^{-1} V^2_\odot
R_\odot$, with $\beta=V_{\rm D}^2/V^2|_{R_\odot} $, i.e.\ the fraction of the disc contribution to
the circular velocity at the Sun. Finally, by exploiting the fact that the quantity
$V/R|_{R_\odot}\equiv \omega = (30.3 \pm 0.3) \ {\rm km/s/kpc}$ is measured with very high accuracy
and much better than $V_\odot$ and $R_\odot$ separately \citep{MB09,reid09},  we obtain
\begin{equation} 
\rho_\odot =  1.2 \times 10^{-27} {\rm
 \frac{ g}{cm^3}}\left(\frac{\omega}{{\rm km/s\,kpc}}\right)^2 \!X_q\bigg[ (1 + 2
{\alpha_\odot}) - 1.1 \,\beta \, f(r_{\odot D})\, X_{z_0} \bigg]\,,
\label{rhoodot}
\end{equation}
where   $r_{\odot D}\equiv R_\odot/R_{\rm D} $ and $f(r_{\odot D})=r_{\odot D}^3H(r_{\odot D})$.

Let us to focus on the advantages of this technique : a) it does not require assuming a particular
DM halo density profile or the dynamical status of some distant tracers of the gravitational field;
b) it is independent of the poorly known values of $V_\odot$ and of the RC slope at different radii
$\alpha(r)$; c) it does not depend on the structural properties of the bulge, which in the mass
modeling, leads to a degeneracy with the values of the mass of stellar disk and of the DM halo. d)
it only mildly depends on the ratio $r_{\odot\rm D}$, as well as, on the disk mass parameter
$\beta$. Let us also note that the determination depends on the RC slope at the Sun
${\alpha_\odot}$, but in a well specified way.

To proceed further we discuss the uncertainties on the parameters appearing in
equation~(\ref{rhoodot}). Our determination depends on the ratio $r_{\odot D}\equiv R_\odot/R_{\rm
  D}$. For this we adopt the reference value and uncertainty $r_{\odot D}= 3.4 \pm 0.5$, as
suggested by the values of $R_D$ discussed above and by the average of values of $R_\odot$ obtained  
recently: $R_\odot=8.2\pm0.5\,$kpc (\cite{ghez,Gillessen:2008qv,bovy}).  This leads to $f(r_{\odot
  D})\simeq 0.42\pm 0.20$, whose uncertainty propagates only mildly into the determination of
$\rho_\odot$, because the second term of the r.h.s.\ of equation~(\ref{rhoodot})\ is only one third
of the first.  

Present data constrain the slope of the circular velocity at the Sun to a central value of
${\alpha_\odot}=0$ and within a fairly narrow range $ -0.075 \leq {\alpha_\odot} \leq 0.075$.  The
uncertainty of $\alpha_\odot$ is the main source of the uncertainty of the present determination of
$\rho_\odot$,  see  for instance~\citep{Olling}.

In equation~(\ref{rhoodot}), $\beta$ is the only quantity that is not observed and therefore
intrinsically uncertain. We can, however, constrain it by computing the maximum value $\beta^M$ for
which the disk contribution at $2.2 \,R_D$ (where it has its maximum) totally accounts for the
circular velocity.  With no assumption on the halo density profile one gets $ \beta^M=0.85$,
independently of $V_\odot$ and $R_\odot$~\citep{PS90}.  However, this is really an absolute maximal
value and it corresponds, out to $R_\odot$, to a solid body halo profile: $V_{h}\propto
R^{\alpha_h}$ with $\alpha_h=1$.  Instead,  mass modeling performed so far for the MW and for
external galaxies tend to find  a lower value: $ \alpha_h (3 R_{\rm D})\leq 0.8$, which yields
$\beta^M=0.77$. We can also set a lower limit for the disk mass, i.e.\ $\beta^m$: first, the
microlensing optical depth to Baade's Window constrains the baryonic matter within the solar circle
to be greater than $3.9\, 10^{10} M_\odot$~\citep{MB09}.  Second, the MW disk B-band luminosity
$L_{\rm B} =2 \times 10 ^{10} L_{\odot}$ and the very reasonable value $M_D/L_B =2 $ again
imply $M_D \simeq 4 \, 10^{10} M_\odot$. All this leads to  $ \beta^m= \beta^M/1.3\simeq 0.65$.  We
thus take $\beta=0.72^{+0.05}_{-0.07}$ as reference range.

Using the reference values and expanding around their central values, we find
\begin{eqnarray}
 \rho_{\odot}\!&=&\!0.43 {\rm \frac{GeV}{cm^3}}\Bigg[1
+ 2.9\,{\alpha_\odot}
-0.64\, \bigg(\beta-0.72\bigg)
+0.45\bigg(r_{\odot D}-3.4\bigg) \nonumber\\
&&{}\qquad\qquad 
- 0.1\left(\frac{z_0}{\text{kpc}}-0.25\right)
+0.10\,\bigg(q-0.95\bigg)+0.07\left(\frac{\omega}{\rm km/s\,kpc}-30.3\right)\Bigg]\,.
\label{eq:10bis}
\end{eqnarray}
This equation estimates the DM density at the Sun's location in an analytic way, in terms of the
involved observational quantities at their present status of knowledge.  The equation is written in
a form such that, for the present reference values of these quantities, the term in the square
brackets on the r.h.s equals 1, and the central result is $\rho_\odot=0.43\,$GeV/cm$^3$. As
such, the determination is ready to account for future improved measurements by simply inserting
them in the r.h.s.\ of~(\ref{eq:10bis}).

The next step is to estimate the uncertainty in the present determination of $\rho_\odot$.  From
equation (\ref{eq:10bis}) and the allowed range of values discussed above, we realize that the main
sources of uncertainty are ${\alpha_\odot}$, $\beta$ and $r_{\odot D}$.  The other parameters give
at most variations of 2-3\%, and can be neglected in the following.

Then, first, it is illustrative to consider ${\alpha_\odot}$, $\beta$ and $r_{\odot D}$ as
independent quantities.  We thus have:
\begin{equation}
 \rho_{\odot}=\bigg(0.43 
\pm 0.094_{({\alpha_\odot})}
\mp 0.016_{(\beta)}
\pm 0.096_{(r_{\odot D})}
\bigg) {\rm \frac{GeV}{cm^3}}
\,,
\label{eq:result2}
\end{equation}
where $A_{(x)}$ means that $A$ is the total effect due to the possible span of the quantity $x$.

At this point, we can go one step further, and assume that the MW is a typical spiral, and
using recent results for the distribution of matter in external galaxies, namely  that DM halos
around spirals are self similar~\citep{URC2} so that the fractional amount of stellar matter $\beta$
{\it directly } dictates  the  value of rotation curve slope ${\alpha_\odot}$~\citep{PS90b}:
\begin{equation}
\beta= 0.72 - 0.95\, {\alpha_\odot}\,.
\end{equation}
Using this relation in equation (\ref{eq:10bis}) we find (neglecting the irrelevant $q$ and $z_0$
terms)
\begin{eqnarray}
 \rho_{\odot}\!&=&\!0.43 {\rm \frac{GeV}{cm^3}}\Bigg[1
+ 3.5\,{\alpha_\odot}
+0.45\bigg(r_{\odot D}-3.4\bigg)+0.07\left(\frac{\omega}{\rm km/s\,kpc}-30.3\right)\Bigg]\,.
\label{eq:11bis}
\end{eqnarray}
From the current known uncertainties, with the estimated range of ${\alpha_\odot}$, we finally
arrive to
\begin{equation}
 \rho_{\odot}=\Big(0.430 
\pm 0.113_{({\alpha_\odot})}
\pm 0.096_{(r_{\odot D})}
\Big) {\rm \frac{GeV}{cm^3}}\,.
\label{eq:result3}
\end{equation} 
 
Its uncertainty mainly reflects our poor knowledge of the velocity slope ${\alpha_\odot}$ and the
uncertainty in the galactocentric Sun distance.

\section{Discussion and conclusion}
We have described here a local determination of $\rho_\odot$, which relies directly on the equation
of centrifugal equilibrium, by estimating the difference between the `total' needed density and that
of the stellar component.  

The method leads to a very reliable kinematical local determination of $\rho_\odot$, avoiding
model-dependent and dubious tasks, mandatory with the standard method, i.e., a) to assume a
particular DM density profile and a specific dynamical status for the tracers of the gravitational
potential, b) to deal with the non-negligible uncertainties of the global MW kinematics, c) to
uniquely disentangle the flattish RC into the different bulge/disk/halo components.

This estimate of $\rho_\odot$ also shows that any kinematical determination using the Galactic
rotation ultimately depends at least on three local quantities, the slope of the circular velocity
at the Sun, the fraction of its amplitude due to the DM, and the ratio between the Sun
galactocentric distance and the disk scale-length, whose uncertainty unavoidably propagates in the
result.  Two of these three quantities can be related by noting that the MW is a typical Spiral and
using the relations available for these kind of galaxies~\citep{URC2}, so that the final uncertainty
can be slightly reduced.

The resulting local DM density that we find is $\rho_{\odot}=(0.43 \pm 0.11_{({\alpha_\odot})}
\pm0.10_{(r_{\odot D})})\, {\rm GeV/cm^{3}}$. We remark that it is free from theoretical assumptions
and can be easily updated by means of equation~(\ref{eq:10bis}) as the relevant observational
quantities will become better known.

Let us comment on other determinations in the literature.  By applying a global modeling method
with a refined statistical analysis to a large set of observational data, \cite{cu09} claimed a very
precise measure, $\rho_\odot=(0.389 \ \pm 0.025)\,$GeV/cm$^{3}$, in agreement with our results but
at variance with our estimated uncertainty (see~\cite{SNG} and also \cite{weber09}). 

 Finally, it is worth discussing the claim by~\cite{Bidin:2012vt}. They, by means of the complex and not
unambiguously accepted technique of exploiting the vertical velocity dispersion of old tracer stars
of the thick disk at the solar neighborhood, claimed the absence of any DM at the solar position, in
disagreement with our (and others) result.  However, as pointed out by~\cite{Bovy:2012tw} their
result is plagued by arbitrary assumptions: different $z$ dependence of the tracers circular
velocity and/or different kinematical status and spatial distribution can trigger a large number of
different possible mass models, some of them clear of DM, others with proportions larger than those
we estimate in~\citep{SNG}.  As a result, the \cite{Bidin:2012vt} method needs a series of
independent checks and increased precision/statistics before being able to deliver a robust
determination of $\rho_\odot$. 

In conclusion, we believe that our technique provides the most trustable estimate of the local dark
matter density and of its uncertainties.

\end{document}